\documentclass[english,aps, prl,  preprint]{revtex4}
\usepackage[T1]{fontenc}
\usepackage[latin9]{inputenc}
\usepackage{color}
\usepackage{babel}
\usepackage{array}
\usepackage{units}
\usepackage{multirow}
\usepackage{amsbsy}
\usepackage{amstext}
\usepackage{amssymb}
\usepackage{graphicx}
\usepackage{esint}
\usepackage[unicode=true,pdfusetitle,
 bookmarks=true,bookmarksnumbered=false,bookmarksopen=false,
 breaklinks=false,pdfborder={0 0 1},backref=false,colorlinks=true]
 {hyperref}
\usepackage{breakurl}

\makeatletter

\providecommand{\tabularnewline}{\\}

\@ifundefined{textcolor}{}
{%
 \definecolor{BLACK}{gray}{0}
 \definecolor{WHITE}{gray}{1}
 \definecolor{RED}{rgb}{1,0,0}
 \definecolor{GREEN}{rgb}{0,1,0}
 \definecolor{BLUE}{rgb}{0,0,1}
 \definecolor{CYAN}{cmyk}{1,0,0,0}
 \definecolor{MAGENTA}{cmyk}{0,1,0,0}
 \definecolor{YELLOW}{cmyk}{0,0,1,0}
}

\makeatother

\begin{document}

\title{Ballistic heat conduction and mass disorder in one dimension}

\author{Zhun-Yong Ong}

\email{ongzy@ihpc.a-star.edu.sg}

\affiliation{Institute of High Performance Computing, A{*}STAR, Singapore 138632}

\author{Gang Zhang}

\email{zhangg@ihpc.a-star.edu.sg}

\affiliation{Institute of High Performance Computing, A{*}STAR, Singapore 138632}
\begin{abstract}
It is well-known that in the disordered harmonic chain, heat conduction
is subballistic and the thermal conductivity ($\kappa$) scales asymptotically
as $\lim_{L\rightarrow\infty}\kappa\propto L^{0.5}$ where $L$ is
the chain length. However, using the nonequilibrium Green's function
(NEGF) method and analytical modeling, we show that there exists a
critical crossover length scale ($L_{C}$) below which ballistic heat
conduction ($\kappa\propto L$) can coexist with mass disorder. This
ballistic-to-subballistic heat conduction crossover is connected to
the exponential attenuation of the phonon transmittance function $\Xi$
\emph{i.e.} $\Xi(\omega,L)=\exp[-L/\lambda(\omega)]$, where $\lambda$
is the frequency-dependent attenuation length. The crossover length
can be determined from the minimum attenuation length which depends
on the maximum transmitted frequency. We numerically determine the
dependence of the transmittance on frequency and mass composition
as well as derive a closed form estimate which agrees closely with
the numerical results. For the length-dependent thermal conductance,
we also derive a closed form expression which agrees closely with
numerical results and reproduces the ballistic to subballistic thermal
conduction crossover. This allows us to characterize the crossover
in terms of changes in the length, mass composition and temperature
dependence, and also to determine the conditions under which heat
conduction enters the ballistic regime. We describe how the mass composition
can be modified to increase ballistic heat conduction
\end{abstract}
\maketitle

\section{Introduction}

It is commonly believed that in the absence of mass disorder, phonons
propagate ballistically without dissipation in pristine one-dimensional
(1D) harmonic systems, resulting in a linear relationship between
the thermal conductivity $\kappa$ and length $L$.\ \cite{SLepri:PhysRep03}
On the other hand, it is well-established from numerical and analytical
results that the presence of mass disorder causes phonon localization\ \cite{HMatsuda:PTP70_Localization,KIshii:PTPS73_Localization,ADhar:PRL01_DisorderedHarmonicChain,BLi_PRL01_CanDisorder,BLi_PRL03_Anomalous,JMDeutsch:PRE03_CorrelationAndScaling,SLepri:PhysRep03,AKundu:PRE10_TimeCorrelation,XNi:PRB11_AnomalousDisordered,JDBodyfelt:PRE13_ScalingTheory},
leading to a sublinear $\kappa$-$L$ relationship ($\kappa\propto L^{\alpha}$
where $\alpha<1$) in the asymptotic ($L\rightarrow\infty$) limit.
This heat conduction scaling behavior is commonly termed subballistic
or superdiffusive in the literature.\ \cite{BLi_PRL01_CanDisorder,BLi_PRL03_Anomalous,SLepri:PhysRep03}
It also implies that the presence of mass disorder and localization\ \cite{PWAnderson:PR58_Localization,KIshii:PTPS73_Localization}
is incompatible with ballistic heat conduction although its mere presence
is not sufficient for the normal heat diffusion behavior.\ \cite{BLi_PRL01_CanDisorder,BLi_PRL03_Anomalous,SLepri:PhysRep03}

However, recent measurements of the thermal conductivity in micrometer-long
SiGe-alloy nanowires at room temperature by Hsiao and co-workers have
yielded a linear $\kappa$-$L$ relationship,\cite{TKHsiao:NatNano13_Observation}
suggesting the possibility of micrometer-scale ballistic phonon transport
despite the presence of mass disorder. Thus, a fresh examination of
the relationship between phonon localization and propagation in 1D
systems is needed to understand this result \emph{i.e.} the coexistence
of ballistic heat conduction and mass disorder. This requires us to
determine the conditions for ballistic conduction to occur in a disordered
1D structure. It also raises the question of how localization varies
with the mass composition of the system and if there are any other
experimentally detectable physical signatures, apart from the linear
$\kappa$-$L$ relationship, that can distinguish ballistic and subballistic
heat conduction. This is important since the experimental determination
of the $\kappa$-$L$ relationship requires the fabrication and measurement
of different samples of varying length,\cite{CWChang:PRL08_Breakdown}
and the process of fabricating samples of different sizes may introduce
variability in the measured thermal conductivity. Such measurement
variability between samples can obscure the signs of ballistic heat
conduction. Therefore, it would be useful to be able to verify this
phenomenon in a \emph{single} sample.

To understand the coexistence of ballistic heat conduction and mass
disorder, we revisit in this paper the phenomenon of phonon propagation
in the disordered 1D harmonic lattice which we take to represent an
idealization of real 1D systems with mass disorder.\cite{ISavic:PRL08_CNTLocalization,TYamamoto:PRL11_Universality,XNi:PRB11_AnomalousDisordered}
Although this model may be simple, it is sufficient for our purpose
of demonstrating and clarifying the coexistence of ballistic heat
conduction and mass disorder. Thus, it can be applied to elucidate
the essential physics of phonon transport in quasi-1D systems with
disorder as well as for the analysis and interpretation of experiments.\ \cite{PKim:PRL01_Thermal,MFujii:PRL05_Measuring,IHsu:APL08,TKHsiao:NatNano13_Observation}
Its simplicity also enables us to relate our numerical results to
known analytical results from conventional wave transmission theory
and to pinpoint the experimentally relevant physical parameters affecting
phonon transmission in the presence of mass disorder. The purpose
of our paper is to analyze the coexistence of ballistic heat conduction
and mass disorder within the established framework of phonon localization.
We show how ballistic heat conduction can occur in the presence of
disorder and how it can be distinguished from subballistic heat conduction.
Although the \emph{asymptotic} scaling theory of heat conduction\ \cite{HMatsuda:PTP70_Localization,ADhar:PRL01_DisorderedHarmonicChain,ISavic:PRL08_CNTLocalization}
and the localization phenomenon are rather well understood, the implications
for \emph{finite} systems have not been clarified. Neither has there
been any attempt to use these disordered 1D models to relate the heat
conduction phenomenon to experimentally relevant variables such as
impurity concentration and temperature.

The organization of our paper is as follows. We first analyze how
mass disorder affects phonon transmitance. While it is known that
mass disorder attenuates phonon propagation and that the transmittance
attenuation length ($\lambda$) scales as the inverse square frequency
($\omega^{-2}$) for low-frequency modes in the $L\rightarrow\infty$
and $\omega\rightarrow0$ limit,\ \cite{HMatsuda:PTP70_Localization,ADhar:PRL01_DisorderedHarmonicChain}
there is no straightforward numerical presentation of this result
in the context of heat conduction. We use the standard nonequilibrium
Green's function (NEGF) method to compute the frequency-dependent
transmittance for chain of different lengths with different values
of impurity mass. We compare the transmittance with known analytical
estimates and find a reasonably good fit over a wide range of frequencies,
allowing us to find the dependence on mass composition. A close analogy
to the Beer-Lambert law is also found. This enables us to find a good
closed form estimate of the thermal conductance, from which we are
able to deduce its dependence on mass composition. We also derive
from the closed form expression the dependence of the thermal conductance
on length, mass composition and temperature. More importantly, we
show and summarize how these dependences differ in the ballistic and
subballistic regime. Given that large lattices are expected to exhibit
a sublinear $\kappa$-$L$ relationship in the $L\rightarrow\infty$
limit, we identify the crossover length scale ($L_{C}$) below which
phonon transport acquires a fully ballistic character independent
of mass composition and above which phonon transport is affected by
lattice disorder. We also show how the crossover length can be tuned
by changing the impurity concentration and mass difference.

\section{Simulation Methodology}

We choose as our model system the familiar 1D harmonic lattice first
studied by Dyson.\ \cite{FJDyson:PR53_DisorderedLinearChain} Our
system differs from that used in other papers in that the atomic mass
is not treated as a continuous random variable. Rather, we have a
random binary lattice (RBL), where the positions of the component
atoms are randomly shuffled, which resembles more realistic systems
with mass disorder and allows us to quantify the effect of experimentally
relevant variables such as impurity concentration and mass difference
on thermal conduction. For simplicity, only adjacent atoms are coupled.
We do not expect our results to be qualitatively different with longer-range
interactions since the interaction typically grows weaker with interatomic
distance. The atoms are only permitted to move longitudinally and
no attempt is made to include any anharmonic interaction in our model. 

\begin{figure}
\includegraphics[width=12cm]{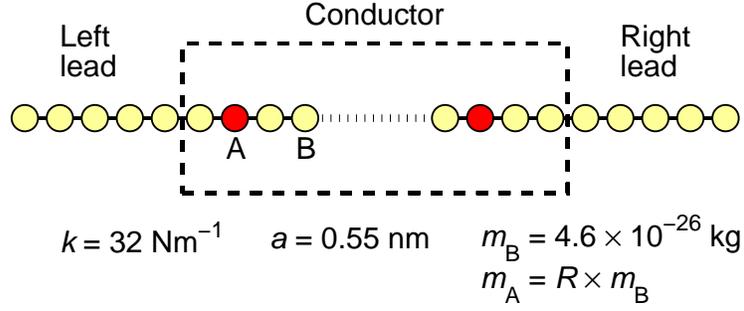}

\caption{Schematic of the the 1D random binary lattice. The atomic parameters
are taken from Ref.\ \cite{WZhang:NHT07}. }
\label{Fig:SimulationSetup}
\end{figure}

Our system consists of three parts. In the middle, there is a finite-size
`conductor' of $N$ equally spaced atoms. On either side, there is
a homogeneous lead \emph{i.e.} a semi-infinite chain with no mass
disorder. In effect, we have a finite disordered system embedded in
an \emph{infinite} homogeneous 1D lattice. Coupling between adjacent
atoms is governed by the harmonic spring term $V(x_{i},x_{i+1})=\frac{1}{2}k(x_{i}-x_{i+1})^{2}$
where $k$ is the spring constant and $x_{i}$ is the displacement
of the $i$-th atom from its equilibrium position. In the RBL, there
are two species of atoms which we label `A' and `B'. Species `A' is
taken to be the substitutional impurity and exists only within the
conductor. The rest of the atoms in the conductor and the leads are
of species `B'. We set the mass of the B atoms to $m_{B}=4.6\times10^{-26}$
kg (the mass of a Si atom), the spring constant to $k=32$ Nm$^{-1}$
(the approximate strength of the Si-Si bond) and the interatomic spacing
to $a=0.55$ nm. Our choice of parameters follows that in Ref.\ \cite{WZhang:NHT07}.
We vary the ratio of the masses ($R=\frac{m_{A}}{m_{B}}$) and concentration
($c_{A}=$ fraction of A atoms, $c_{B}=1-c_{A}$) in our simulations.
The 1D lattice size is varied from $N=100$ to 20000, and the equivalent
range of chain length values ($L=Na$) is $L=55$ nm to $1.1$ $\mu$m.
A schematic of the system is shown in Fig.\ \ref{Fig:SimulationSetup}.

To simulate phonon transport, we use the nonequilibrium Green's function
(NEGF) method which is described in detail in Refs.\ \cite{WZhang:NHT07,NMingo:Springer09}.
The method calculates the frequency-dependent energy transmittance
function $\Xi(\omega)$ by modeling the conductor as an open quantum
system and the semi-infinite leads as self-energies (open boundaries).
The Green's function for the conductor is 
\[
\boldsymbol{G}(\omega)=\left[\omega^{2}+i0^{+}-\boldsymbol{K_{C}}-\boldsymbol{\Sigma_{L}}(\omega)-\boldsymbol{\Sigma_{R}}(\omega)\right]^{-1}\ ,
\]
where $\boldsymbol{\Sigma_{L}}$ ($\boldsymbol{\Sigma_{R}}$) is the
so-called self-energy corresponding to the left (right) semi-infinite
lead; $\boldsymbol{K_{C}}$ is the force constant matrix of the conductor,
and its $(i,j)$ matrix element which couples the $i$-th and $j$-th
atom is given by 
\[
[\boldsymbol{K_{C}}]_{ij}=(m_{i}m_{j})^{-\nicefrac{1}{2}}\frac{\partial^{2}V}{\partial x_{i}\partial x_{j}}\ .
\]
To compute $\Xi(\omega)$, we use the Caroli transmission formula\ \cite{CCaroli:JPhysC71}
$\Xi(\omega)=\text{Tr}(\boldsymbol{\Gamma_{L}G\Gamma_{R}G^{\dagger}})$,
where $\boldsymbol{\Gamma_{L}}$ ($\boldsymbol{\Gamma_{R}}$) is the
term coupling the conductor to left (right) lead.\ \cite{WZhang:NHT07,NMingo:Springer09}
As with any study of disordered systems, it is necessary to express
the computed variables in terms of their ensemble average. In this
paper, the ensemble-average transmittance function $\langle\Xi(\omega)\rangle$
is obtained by averaging $\Xi(\omega)$ over 100 realizations of the
RBL.

It is more useful to work with the thermal conductance $\sigma$ which
is related to the thermal conductivity via the relation $\kappa\propto\sigma L$.
The thermal conductance is computed using the Landauer formula:\ \cite{WZhang:NHT07,NMingo:Springer09}
\begin{equation}
\sigma(T)=\frac{1}{2\pi}\int_{0}^{\infty}\hbar\omega\frac{df(\omega)}{dT}\langle\Xi(\omega)\rangle\ d\omega\label{Eq:LandauerFormula}
\end{equation}
where $T$ and $\hbar$ are respectively the temperature and reduced
Planck constant, and $f(\omega)=[\exp(\frac{\hbar\omega}{k_{B}T})-1]^{-1}$is
the Bose-Einstein occupation factor. The expression in Eq.\ (\ref{Eq:LandauerFormula})
integrates the frequency-dependent transmittance spectrum and the
differential heat capacity over the entire frequency range ($0$ to
$\infty$). In the high temperature limit, we have $f(\omega)\approx\frac{k_{B}T}{\hbar\omega}$
so that Eq.\ (\ref{Eq:LandauerFormula}) becomes:
\begin{equation}
\lim_{T\rightarrow\infty}\sigma(T)=\frac{k_{B}}{2\pi}\int_{0}^{\infty}\langle\Xi(\omega)\rangle\ d\omega\ ,\label{Eq:ClassicalLandauerFormula}
\end{equation}
which is proportional to the \emph{total} transmittance. In the case
of the homogeneous chain ($m_{A}=m_{B}$), the transmittance is $\Xi(\omega)=\Theta(\omega_{C}-\omega)$
where $\omega_{C}=2\sqrt{k/m_{B}}$ is the frequency cutoff, and Eq.\ (\ref{Eq:ClassicalLandauerFormula})
yields a length-independent thermal conductance \emph{i.e.} $\lim_{T\rightarrow\infty}\sigma(T)=\frac{k_{B}\omega{}_{C}}{2\pi}$.
This example implies that a length-dependent transmittance function
is needed for the conductance to vary with $L$.

\section{Results and Discussion}

\subsection{Length dependence of transmittance spectrum}

\begin{figure}
\includegraphics[width=12cm]{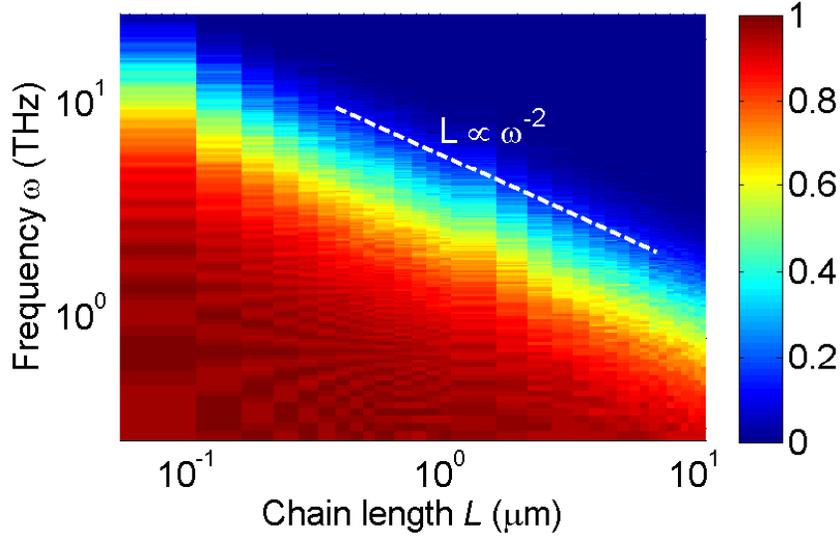}

\caption{Two-dimensional plot of the transmittance function $\langle\Xi(\omega,L)\rangle$
as a function of frequency $\omega$ and chain length $L$ for $c_{A}=0.5$
and $R=2$. The plot shows that the transmittance decreases as the
chain length increases. The transmittance of the low-frequency modes
is relatively unattenuated by the increase in $L$ while the transmittance
of the higher-frequency modes ($\omega\gtrsim1$ THz) decays rapidly
with increasing $L$. The position of the `shoulder' ($\omega$),
corresponding to the cyan region and $\langle\Xi\rangle\sim1/e$,
shifts towards the origin as $L$ increases and varies as $L\propto\omega^{-2}$
(line is drawn in white). }

\label{Fig:TransmittanceLengthScaling}
\end{figure}

Given that the transmittance spectrum is expected to vary with $L$,
the question arises as to what its dependence on $L$ should be. In
general, the transmittance function can be interpreted as the ease
with which a phonon traverses the conductor. To identify the functional
relationship, we first compute $\langle\Xi(\omega)\rangle$ for different
values of chain length $L$ at different values of $R$ and $c_{A}$.
Figure\ \ref{Fig:TransmittanceLengthScaling} shows $\langle\Xi(\omega)\rangle$
decreasing with increasing $\omega$ and $L$ for a particular combination
of $c_{A}$ and $R$ ($c_{A}=0.5$ and $R=2$), with the rate of decrease
faster for higher-frequency modes. From our calculations, we find
that $\langle\Xi(\omega,L)\rangle$ obeys the simple numerical relation:
\begin{equation}
\langle\Xi(\omega,L)\rangle=\langle\Xi(\omega,L_{0})\rangle\exp\left[-\frac{L-L_{0}}{\lambda(\omega)}\right]\ ,\label{Eq:BeerLambertLaw}
\end{equation}
for $\omega<\omega_{C}$ and $L>L_{0}$, analogous to the \emph{Beer-Lambert}
law in optics. The parameter $\lambda(\omega)$ is the frequency-dependent
attenuation length and can be interpreted as the distance through
which the phonon has its transmitted square amplitude reduced by a
factor of $1/e$. Numerical fitting yields the numerical relationship
$\lambda\propto\omega^{-2}$. 

\begin{figure}
\includegraphics[width=12cm]{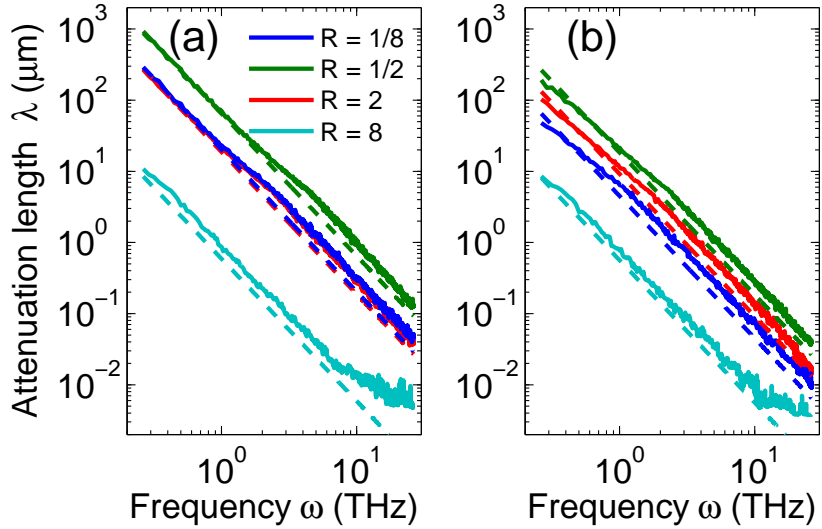}

\caption{Plot of the extracted attenuation length $\lambda$ (solid lines)
and its low-frequency estimate $\lambda_{0}$ (dashed lines) as a
function of $\omega$ for $R$ = 1/8, 1/2, 2 and 8 at (a) $c_{A}=0.1$
and (b) $c_{A}=0.5$. At low frequency ($\omega<1$ THz), there is
very good agreement between $\lambda_{0}$ and $\lambda$, as expected.
At higher frequencies, $\lambda$ is slightly larger than $\lambda_{0}$.
The approximation $\lambda_{0}$ however diverges increasingly from
the extracted attenuation length $\lambda$ as $\lambda_{0}$ approaches
$a$, the interatomic spacing. }
\label{Fig:AttenuationLengthFit}
\end{figure}

To determine the exact relationship between $\lambda$ and $\omega$,
we use the recent result from Das and Dhar,\ \cite{SGDas:EPJB12_Landauer}
who proved an intuitive and useful formula relating the energy transmittance
$\Xi(\omega)$ and the plane-wave transmission function $\tau(\omega)$
for the 1D harmonic chain:
\begin{equation}
\Xi(\omega)=|\tau(\omega)|^{2}\ .\label{Eq:DasDharResult}
\end{equation}
In the disordered harmonic chain with $N$ atoms, the ensemble average
of its asymptotic amplitude has the form of an exponentially decaying
function \emph{i.e.} $\lim_{N\rightarrow\infty}|\tau(\omega)|=\exp[-\gamma(\omega)N]$
where $\gamma$ is the dimensionless Lyapunov exponent.\ \cite{HMatsuda:PTP70_Localization,ADhar:PRL01_DisorderedHarmonicChain}
This implies that the attenuation length in Eq.\ (\ref{Eq:BeerLambertLaw})
is 
\begin{equation}
\lim_{L\rightarrow\infty}\lambda(\omega)=\frac{a}{2\gamma(\omega)}\ .\label{Eq:AttenuationLengthResult}
\end{equation}
In the low frequency ($\omega\rightarrow0^{+}$) limit, the closed
form expression for $\gamma$ is known\ \cite{HMatsuda:PTP70_Localization}
and given by 
\begin{equation}
\lim_{\omega\rightarrow0}\gamma(\omega)=\frac{\langle\delta m\rangle^{2}\omega^{2}}{8k\langle m\rangle}\ ,\label{Eq:LowFreqInvLocalLength}
\end{equation}
where $\langle\delta m\rangle$ and $\langle m\rangle$ are respectively
the standard deviation and the mean of the conductor atomic mass.
Combining Eqs.\ (\ref{Eq:LowFreqInvLocalLength}) and (\ref{Eq:AttenuationLengthResult}),
we obtain the expression for the low-frequency attenuation length
\begin{equation}
\lambda_{0}(\omega)=\frac{4ka\langle m\rangle}{\langle\delta m\rangle^{2}\omega^{2}}\label{Eq:LowFreqAttenLength}
\end{equation}
which we plot in Fig.\ \ref{Fig:AttenuationLengthFit} together with
$\lambda(\omega)$ for $R$ = 1/8, 1/2, 2 and 8 at $c_{A}=0.1$ and
$0.5$. We find close agreement between the two at low frequencies
($\omega<1$ THz). At higher frequencies, $\lambda_{0}(\omega)$ is
slightly smaller but still an excellent approximation to the numerically
computed $\lambda(\omega)$. The approximation $\lambda_{0}$ however
diverges increasingly from the extracted attenuation length $\lambda$
as $\lambda_{0}$ approaches $a$, the interatomic spacing. This is
more evident for the $R=8$ curve in Fig.\ \ref{Fig:AttenuationLengthFit}
when the impurity mass $m_{A}$ is much greater than $m_{B}$. 

If we interpret $\lambda_{0}(\omega)$ as an impurity-limited mean
free path (MFP) and assume a very dilute impurity concentration $n_{\text{imp}}=c_{A}\ll1$,
then we have $\langle\delta m\rangle^{2}=c_{A}(1-c_{A})(m_{A}-m_{B})^{2}\approx n_{\text{imp}}(m_{A}-m_{B})^{2}$
and obtain $\lambda_{0}\propto n_{\text{imp}}^{-1}(m_{A}-m_{B})^{-2}\omega^{-2}$
\emph{i.e.} the attenuation length varies as $n_{\text{imp}}^{-1}$.
This parallels the Beer-Lambert law for photons where the attenuation
length is inversely proportional to the absorber concentration. In
three dimensions, the impurity-limited MFP {[}$\lambda_{\text{imp,3D}}\propto n_{\text{imp}}^{-1}(m_{A}-m_{B})^{-2}\omega^{-4}${]}
has identical dependence on impurity concentration and mass difference
but a very different frequency dependence.\ \cite{MGHolland:PR63_Analysis} 

\subsection{Length, mass composition and temperature dependence for ballistic
and subballistic thermal conductance}

Having numerically verified the form of the transmittance function,
we proceed to calculate the thermal conductance to determine its dependence
on length, mass composition and temperature. From Eqs.\ (\ref{Eq:LandauerFormula}),
(\ref{Eq:LowFreqInvLocalLength}) and (\ref{Eq:AttenuationLengthResult}),
we arrive at an analytical expression for estimating the length-dependent
thermal conductance:
\begin{equation}
\sigma_{0}(T,L)=\frac{1}{2\pi}\int_{0}^{\omega_{C}}\hbar\omega\frac{df}{dT}\exp\left[-\frac{L\langle\delta m\rangle^{2}\omega^{2}}{4ka\langle m\rangle}\right]\ d\omega\ .\label{Eq:AnalyticalConductanceFormula}
\end{equation}
This approximation relies on the $\lim_{L\rightarrow\infty,\omega\rightarrow0^{+}}\lambda(\omega)$
expression in Eq.\ (\ref{Eq:AttenuationLengthResult})\emph{ i.e.}
it makes use of the \emph{low-frequency} approximation for $\lambda(\omega)$
in the $L\rightarrow\infty$ limit. At finite $L$, higher-frequency
modes contribute more to the phonon transmittance and the their attenuation
behavior may not follow that of Eq.\ (\ref{Eq:LowFreqInvLocalLength}).

\begin{figure}
\includegraphics[width=12cm]{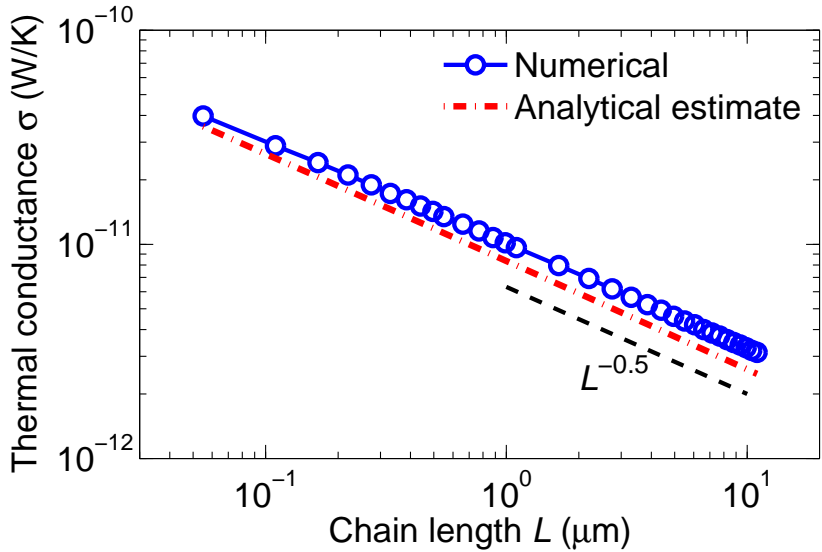}

\caption{Plot of the length-dependent thermal conductances $\sigma$ (circle)
and $\sigma_{0}$ (dash-dot) computed respectively from the NEGF method
and Eq.\ (\ref{Eq:AnalyticalConductanceFormula}) for $R=2$ and
$c_{A}=0.1$ at $T=300$ K. $\sigma_{0}$ scales as $L^{-0.5}$ but
is about 20 percent smaller than $\sigma$. }
\label{Fig:ConductanceFormula}
\end{figure}

To see how well Eq.\ (\ref{Eq:AnalyticalConductanceFormula}) approximates
the thermal conductance $\sigma$ (calculated from NEGF) especially
for \emph{finite} $L$, we compute $\sigma_{0}$ for $R=2$ and $c_{A}=0.1$
at $T=300$ K and plot it in Fig.\ \ref{Fig:ConductanceFormula}
together with the corresponding $\sigma$. $\sigma_{0}$ scales as
$L^{-0.5}$ and agrees well with $\sigma$ but is consistently $\sim20$
percent smaller. We have also computed $\sigma_{0}$ for other values
of $R$ and $c_{A}$ and found that it is also $\sim20$ percent smaller
than $\sigma$ computed from NEGF. This small discrepancy is expected
given that Eq.\ (\ref{Eq:AnalyticalConductanceFormula}) uses the
low-frequency approximation $\lambda_{0}$ which overestimates the
transmittance attenuation. 

Since Fig.\ (\ref{Fig:ConductanceFormula}) shows that the expression
for $\sigma_{0}$ from Eq.\ (\ref{Eq:AnalyticalConductanceFormula})
can fit the numerically computed $\sigma$, it means that we can use
Eq.\ (\ref{Eq:AnalyticalConductanceFormula}) to infer the asymptotic
behavior of the conductance in the $L\rightarrow\infty$ and $L\rightarrow0$
limits. In the $L\rightarrow\infty$ limit, only very low frequency
modes contribute to the integral in Eq.\ (\ref{Eq:AnalyticalConductanceFormula}),
allowing us to use the high-temperature approximation \emph{i.e.}
\[
\frac{df}{dT}=\frac{k_{B}}{\hbar\omega}\ ,
\]
so that Eq.\ (\ref{Eq:AnalyticalConductanceFormula}) becomes 
\[
\lim_{L\rightarrow\infty}\sigma_{0}(T,L)\approx\frac{k_{B}}{2\pi}\int_{0}^{\infty}\exp\left[-\frac{L\langle\delta m\rangle^{2}\omega^{2}}{4ka\langle m\rangle}\right]\ d\omega\ .
\]
This yields 
\begin{equation}
\lim_{L\rightarrow\infty}\sigma_{0}=\left[\frac{ka\langle m\rangle k_{B}^{2}}{4\pi\langle\delta m\rangle^{2}}\right]^{\frac{1}{2}}\frac{1}{\sqrt{L}}\label{Eq:SuperDiffusiveConductance}
\end{equation}
which gives the well-known $\lim_{L\rightarrow\infty}\sigma\propto L^{-0.5}$
relationship,\ \cite{HMatsuda:PTP70_Localization,ADhar:PRL01_DisorderedHarmonicChain,XNi:PRB11_AnomalousDisordered}
indicating subballistic (superdiffusive) thermal transport. In the
$L\rightarrow0$ limit, the conductor is transparent to all phonon
modes with frequency below $\omega_{C}$ (the cutoff frequency determined
by the atomic mass in the leads, $m_{B}$), \emph{i.e.} $\Xi(\omega)\approx\Theta(\omega_{C}-\omega)$,
and we recover the expression for the finite length-independent thermal
conductance, 
\begin{equation}
\lim_{L\rightarrow0}\sigma_{0}=\frac{1}{2\pi}\int_{0}^{\omega_{C}}\hbar\omega\frac{df}{dT}\ d\omega\label{Eq:BallisticConductance}
\end{equation}
which is \emph{independent} of the mass composition. We note that
the ballistic conductance in Eq.\ (\ref{Eq:BallisticConductance})
is proportional to the heat capacity of the system.

Although the model described by Eq.\ (\ref{Eq:AnalyticalConductanceFormula})
is very simple, the variation in its thermal transport character in
the different limits, as suggested by Eqs. (\ref{Eq:SuperDiffusiveConductance})
and (\ref{Eq:BallisticConductance}), has significant qualitative
implications for more realistic experimental systems, apart from its
$\sigma\propto L^{-0.5}$ property. We use them to deduce the difference
in the dependence on the various physical variables. Firstly, in the
\emph{ballistic} limit, the thermal conductance is \emph{independent}
of its mass composition. On the other hand, in the \emph{subballistic}
limit, the conductance is inversely proportional to the standard deviation
of the mass concentration, and in the \emph{dilute} limit ($c_{A}\rightarrow0$),
it should scale as $\sigma\propto n_{\text{imp}}^{-0.5}|\Delta m|^{-1}$,
where $\Delta m=m_{A}-m_{B}$ is the mass difference, since $\langle\delta m\rangle^{2}=c_{A}c_{B}(m_{A}-m_{B})^{2}\approx n_{\text{imp}}\Delta m{}^{2}$
in Eq.\ (\ref{Eq:SuperDiffusiveConductance}). Secondly, the temperature
dependence of the conductance is different for the subballistic and
ballistic case. In the subballistic case ($L\rightarrow\infty$),
the range of conducting phonon modes decreases as $L^{-0.5}$ and
only the low-frequency modes contribute to thermal transport. Hence,
the temperature dependence of the thermal conductance is negligible.
However, in the ballistic case ($L\rightarrow0$), the range of conducting
modes ($\omega<\omega_{C}$) is determined by the phonon occupation
in the leads. Thus, the conductance is more sensitive to temperature
change and in the low-temperature limit ($T\ll\hbar\omega_{C}/k_{B}$),
we obtain $\sigma_{0}\propto T$. We summarize the differences in
the scaling dependence of the thermal conductance in Table \ref{Tab:ScalingRelationships}.
These differences can be used to distinguish the ballistic and subballistic
transport regimes.

\begin{table}
\begin{tabular}{|l|cc|}
\hline 
\multirow{3}{*}{Physical variable} & \multicolumn{2}{c|}{Scaling exponent $\alpha$}\tabularnewline
\cline{2-3} 
 & Ballistic & Subballistic\tabularnewline
 & ($L\ll L_{C}$) & ($L\gg L_{C}$)\tabularnewline
\hline 
\hline 
Length ($\sigma\propto L^{\alpha}$) & 0 & -0.5 to -0.4\tabularnewline
Temperature ($\sigma\propto T^{\alpha}$) & 1 & 0\tabularnewline
Mass difference ($\sigma\propto|\Delta m|^{\alpha}$) & 0 & -1\tabularnewline
Impurity concentration ($\sigma\propto n_{\text{imp}}^{\alpha}$) & 0 & -0.5\tabularnewline
\hline 
\end{tabular}

\caption{Scaling dependence of the thermal conductance $\sigma$ on various
physical variables (length, temperature, mass difference and impurity
concentration) in the ballistic ($L\ll L_{C}$) and subballistic ($L\gg L_{C}$)
regime in the 1D random binary lattice. }

\label{Tab:ScalingRelationships}
\end{table}

\subsection{Ballistic to subballistic crossover length}

Given that thermal conduction is ballistic in the $L\rightarrow0$
limit and subballistic in the $L\rightarrow\infty$ limit, there must
exist an intermediate length scale that marks the crossover from ballistic
to subballistic thermal transport although we do not observe one in
Fig.\ \ref{Fig:ConductanceFormula}. Since the thermal conductance
depends on the mass composition in the $L\rightarrow\infty$ limit
{[}Eq.\ (\ref{Eq:SuperDiffusiveConductance}){]}, it implies that
the crossover length varies with mass composition. This can be increased
by weakening the amount of mass disorder so that ballistic heat conduction
is detectable in the range of $L$ values simulated.

\begin{figure}
\includegraphics[width=12cm]{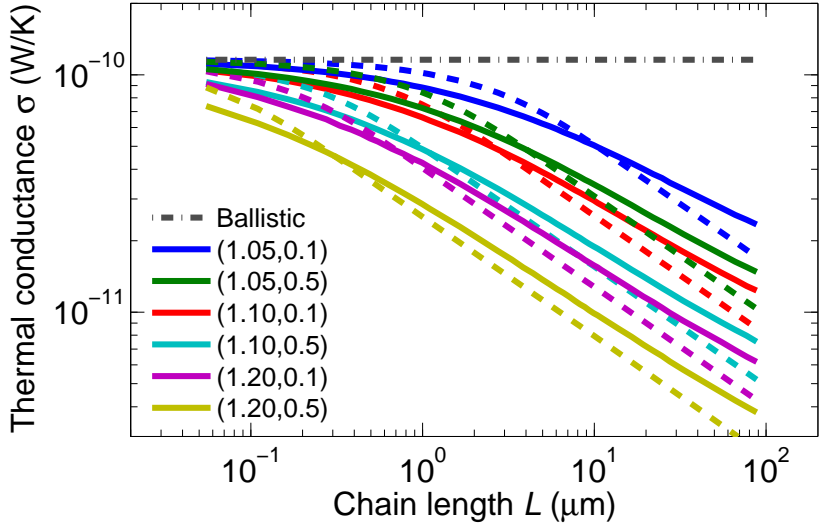}

\caption{Plot of the NEGF-computed $\sigma$ (solid line) and $\sigma_{0}$
from Eq.\ (\ref{Eq:AnalyticalConductanceFormula}) (dashed line)
for $(R,c_{A})$ equal to $(1.05,0.1)$, $(1.05,0.5)$, $(1.1,0.1)$,
$(1.1,0.5)$, $(1.2,0.1)$ and $(1.2,0.5)$ in the high temperature
($T\rightarrow\infty$) limit in blue, green, red, cyan, magenta and
yellow, respectively. The dash-dotted line shows the ballistic conductance
in the homogeneous chain. There is good agreement between $\sigma$
and $\sigma_{0}$, with the fit improving as $R$ increases. The ballistic-to-subballistic
crossover is around $L_{C}$, ranging between 0.05 to 2 $\mu$m. }

\label{Fig:CrossoverConductance}
\end{figure}

For a dilute impurity concentration and small mass difference, the
mass variance ($\langle\delta m\rangle^{2}$) is small and the mass
disorder is weak. Thus, the crossover length between ballistic and
subballistic thermal transport can be large. The crossover length
scale $L_{C}$ can be estimated by estimating the minimum possible
attenuation length which is determined by the cutoff frequency $\omega_{C}$.
Using Eq.\ (\ref{Eq:LowFreqAttenLength}), we set $L_{C}=\lambda_{0}(\omega_{C})$
which yields $L_{C}=a\langle m\rangle m_{B}/[c_{A}(1-c_{A})\Delta m^{2}]$.
This suggests that the ballistic regime can be extended by reducing
the impurity concentration as well as mass ratio ($R$). Instead of
using $R=2$, we pick a small $R$ so that the mass difference would
be small ($\Delta m\ll m_{B}$). Figure\ \ref{Fig:CrossoverConductance}
shows the NEGF-computed $\sigma$ and $\sigma_{0}$ from Eq.\ (\ref{Eq:AnalyticalConductanceFormula})
as a function of $L$ (56 nm to 88 $\mu$m) for $(R,C_{A})$ equal
to $(1.05,0.1)$, $(1.05,0.5)$, $(1.1,0.1)$, $(1.1,0.5)$, $(1.2,0.1)$
and $(1.2,0.5)$ in the high temperature ($T\rightarrow\infty$) limit.
We find good agreement between $\sigma$ and $\sigma_{0}$ over the
range of $L$ considered, with the agreement better for larger $c_{A}$
and $\Delta m$. The corresponding crossover length ranges between
$L_{C}\approx0.05$ to 2 $\mu$m. Below $L<L_{C}$, $\sigma$ and
$\sigma_{0}$ converge to the ballistic limit $\sigma=\frac{k_{B}\omega{}_{C}}{2\pi}$
while for $L\gg L_{C}$, $\sigma\propto L^{-\alpha}$ with $\alpha<0.5$
and $\sigma_{0}\propto L^{-0.5}$, respectively. The discrepancy in
the length dependence between $\sigma$ and $\sigma_{0}$ may be a
consequence of the small mass variance. In the $\langle\delta m\rangle\rightarrow0$
limit, we expect fully ballistic thermal transport\emph{ i.e.} $\sigma\propto L^{0}$.
Thus, the scaling exponent would deviate from -0.5 toward 0 for small
$\langle\delta m\rangle$. 

\section{Summary}

In summary, we have studied heat conduction in the 1D random binary
lattice model with mass disorder. We have shown that the energy transmittance
attenuates exponentially with lattice size, similar to the Beer-Lambert
law for photons. The attenuation length is shown numerically to have
frequency dependence of $\lambda\propto\omega^{-2}$ for low frequency
modes (as predicted by localization theory) and also for high frequency
modes (not predicted by localization theory). The existence of a maximum
cutoff frequency implies a minimum attenuation length below which
the transmittance of the phonons in the system converges to unity.
This indicates that ballistic heat conduction can coexist with mass
disorder for finite systems. The crossover length between ballistic
and subballistic heat conduction can be estimated based on the cutoff
frequency $\omega_{c}$ and the low-frequency expression for the attenuation
length $\lambda_{0}$. This analytical estimate has been shown to
agree closely with the numerical crossover length computed with NEGF.
We have also derived the explicit length, temperature and mass composition
dependence of the thermal conductance and shown how they differ in
the ballistic and subballistic limit. The differences can be used
to distinguish ballistic and subballistic heat conduction in experimental
systems. By making use of the mass composition dependence of the attenuation
length, we have shown how the crossover between ballistic and subballistic
heat conduction can be manipulated by changing the impurity mass difference
and concentration. 

The authors gratefully acknowledge the financial support from the
Agency for Science, Technology and Research (A{*}STAR), Singapore
and the use of computing resources at the A{*}STAR Computational Resource
Centre, Singapore.

\appendix

\bibliographystyle{apsrev4-1}
\bibliography{PhononNotes}

\begin{thebibliography}{24}%
\makeatletter
\providecommand \@ifxundefined [1]{%
 \@ifx{#1\undefined}
}%
\providecommand \@ifnum [1]{%
 \ifnum #1\expandafter \@firstoftwo
 \else \expandafter \@secondoftwo
 \fi
}%
\providecommand \@ifx [1]{%
 \ifx #1\expandafter \@firstoftwo
 \else \expandafter \@secondoftwo
 \fi
}%
\providecommand \natexlab [1]{#1}%
\providecommand \enquote  [1]{``#1''}%
\providecommand \bibnamefont  [1]{#1}%
\providecommand \bibfnamefont [1]{#1}%
\providecommand \citenamefont [1]{#1}%
\providecommand \href@noop [0]{\@secondoftwo}%
\providecommand \href [0]{\begingroup \@sanitize@url \@href}%
\providecommand \@href[1]{\@@startlink{#1}\@@href}%
\providecommand \@@href[1]{\endgroup#1\@@endlink}%
\providecommand \@sanitize@url [0]{\catcode `\\12\catcode `\$12\catcode
  `\&12\catcode `\#12\catcode `\^12\catcode `\_12\catcode `\%12\relax}%
\providecommand \@@startlink[1]{}%
\providecommand \@@endlink[0]{}%
\providecommand \url  [0]{\begingroup\@sanitize@url \@url }%
\providecommand \@url [1]{\endgroup\@href {#1}{\urlprefix }}%
\providecommand \urlprefix  [0]{URL }%
\providecommand \Eprint [0]{\href }%
\providecommand \doibase [0]{http://dx.doi.org/}%
\providecommand \selectlanguage [0]{\@gobble}%
\providecommand \bibinfo  [0]{\@secondoftwo}%
\providecommand \bibfield  [0]{\@secondoftwo}%
\providecommand \translation [1]{[#1]}%
\providecommand \BibitemOpen [0]{}%
\providecommand \bibitemStop [0]{}%
\providecommand \bibitemNoStop [0]{.\EOS\space}%
\providecommand \EOS [0]{\spacefactor3000\relax}%
\providecommand \BibitemShut  [1]{\csname bibitem#1\endcsname}%
\let\auto@bib@innerbib\@empty
\bibitem [{\citenamefont {Lepri}\ \emph {et~al.}(2003)\citenamefont {Lepri},
  \citenamefont {Livi},\ and\ \citenamefont {Politi}}]{SLepri:PhysRep03}%
  \BibitemOpen
  \bibfield  {author} {\bibinfo {author} {\bibfnamefont {S.}~\bibnamefont
  {Lepri}}, \bibinfo {author} {\bibfnamefont {R.}~\bibnamefont {Livi}}, \ and\
  \bibinfo {author} {\bibfnamefont {A.}~\bibnamefont {Politi}},\ }\href@noop {}
  {\bibfield  {journal} {\bibinfo  {journal} {Physics Reports}\ }\textbf
  {\bibinfo {volume} {377}},\ \bibinfo {pages} {1} (\bibinfo {year}
  {2003})}\BibitemShut {NoStop}%
\bibitem [{\citenamefont {Matsuda}\ and\ \citenamefont
  {Ishii}(1970)}]{HMatsuda:PTP70_Localization}%
  \BibitemOpen
  \bibfield  {author} {\bibinfo {author} {\bibfnamefont {H.}~\bibnamefont
  {Matsuda}}\ and\ \bibinfo {author} {\bibfnamefont {K.}~\bibnamefont
  {Ishii}},\ }\href@noop {} {\bibfield  {journal} {\bibinfo  {journal} {Suppl.
  Prog. Theor. Phys.}\ }\textbf {\bibinfo {volume} {45}},\ \bibinfo {pages}
  {56} (\bibinfo {year} {1970})}\BibitemShut {NoStop}%
\bibitem [{\citenamefont {Ishii}(1973)}]{KIshii:PTPS73_Localization}%
  \BibitemOpen
  \bibfield  {author} {\bibinfo {author} {\bibfnamefont {K.}~\bibnamefont
  {Ishii}},\ }\href@noop {} {\bibfield  {journal} {\bibinfo  {journal} {Suppl.
  Prog. Theor. Phys.}\ }\textbf {\bibinfo {volume} {53}},\ \bibinfo {pages}
  {77} (\bibinfo {year} {1973})}\BibitemShut {NoStop}%
\bibitem [{\citenamefont {Dhar}(2001)}]{ADhar:PRL01_DisorderedHarmonicChain}%
  \BibitemOpen
  \bibfield  {author} {\bibinfo {author} {\bibfnamefont {A.}~\bibnamefont
  {Dhar}},\ }\href@noop {} {\bibfield  {journal} {\bibinfo  {journal} {Phys.
  Rev. Lett.}\ }\textbf {\bibinfo {volume} {86}},\ \bibinfo {pages} {5882}
  (\bibinfo {year} {2001})}\BibitemShut {NoStop}%
\bibitem [{\citenamefont {Li}\ \emph {et~al.}(2001)\citenamefont {Li},
  \citenamefont {Zhao},\ and\ \citenamefont {Hu}}]{BLi_PRL01_CanDisorder}%
  \BibitemOpen
  \bibfield  {author} {\bibinfo {author} {\bibfnamefont {B.}~\bibnamefont
  {Li}}, \bibinfo {author} {\bibfnamefont {H.}~\bibnamefont {Zhao}}, \ and\
  \bibinfo {author} {\bibfnamefont {B.}~\bibnamefont {Hu}},\ }\href@noop {}
  {\bibfield  {journal} {\bibinfo  {journal} {Phys. Rev. Lett.}\ }\textbf
  {\bibinfo {volume} {86}},\ \bibinfo {pages} {63} (\bibinfo {year}
  {2001})}\BibitemShut {NoStop}%
\bibitem [{\citenamefont {Li}\ and\ \citenamefont
  {Wang}(2003)}]{BLi_PRL03_Anomalous}%
  \BibitemOpen
  \bibfield  {author} {\bibinfo {author} {\bibfnamefont {B.}~\bibnamefont
  {Li}}\ and\ \bibinfo {author} {\bibfnamefont {J.}~\bibnamefont {Wang}},\
  }\href@noop {} {\bibfield  {journal} {\bibinfo  {journal} {Phys. Rev. Lett.}\
  }\textbf {\bibinfo {volume} {91}},\ \bibinfo {pages} {044301} (\bibinfo
  {year} {2003})}\BibitemShut {NoStop}%
\bibitem [{\citenamefont {Deutsch}\ and\ \citenamefont
  {Narayan}(2003)}]{JMDeutsch:PRE03_CorrelationAndScaling}%
  \BibitemOpen
  \bibfield  {author} {\bibinfo {author} {\bibfnamefont {J.~M.}\ \bibnamefont
  {Deutsch}}\ and\ \bibinfo {author} {\bibfnamefont {O.}~\bibnamefont
  {Narayan}},\ }\href@noop {} {\bibfield  {journal} {\bibinfo  {journal} {Phys.
  Rev. E}\ }\textbf {\bibinfo {volume} {68}},\ \bibinfo {pages} {041203}
  (\bibinfo {year} {2003})}\BibitemShut {NoStop}%
\bibitem [{\citenamefont {Kundu}(2010)}]{AKundu:PRE10_TimeCorrelation}%
  \BibitemOpen
  \bibfield  {author} {\bibinfo {author} {\bibfnamefont {A.}~\bibnamefont
  {Kundu}},\ }\href@noop {} {\bibfield  {journal} {\bibinfo  {journal} {Phys.
  Rev. E}\ }\textbf {\bibinfo {volume} {82}},\ \bibinfo {pages} {031131}
  (\bibinfo {year} {2010})}\BibitemShut {NoStop}%
\bibitem [{\citenamefont {Ni}\ \emph {et~al.}(2011)\citenamefont {Ni},
  \citenamefont {Leek}, \citenamefont {Wang}, \citenamefont {Feng},\ and\
  \citenamefont {Li}}]{XNi:PRB11_AnomalousDisordered}%
  \BibitemOpen
  \bibfield  {author} {\bibinfo {author} {\bibfnamefont {X.}~\bibnamefont
  {Ni}}, \bibinfo {author} {\bibfnamefont {M.~L.}\ \bibnamefont {Leek}},
  \bibinfo {author} {\bibfnamefont {J.-S.}\ \bibnamefont {Wang}}, \bibinfo
  {author} {\bibfnamefont {Y.~P.}\ \bibnamefont {Feng}}, \ and\ \bibinfo
  {author} {\bibfnamefont {B.}~\bibnamefont {Li}},\ }\href@noop {} {\bibfield
  {journal} {\bibinfo  {journal} {Phys. Rev. B}\ }\textbf {\bibinfo {volume}
  {83}},\ \bibinfo {pages} {045408} (\bibinfo {year} {2011})}\BibitemShut
  {NoStop}%
\bibitem [{\citenamefont {Bodyfelt}\ \emph {et~al.}(2013)\citenamefont
  {Bodyfelt}, \citenamefont {Zheng}, \citenamefont {Fleischmann},\ and\
  \citenamefont {Kottos}}]{JDBodyfelt:PRE13_ScalingTheory}%
  \BibitemOpen
  \bibfield  {author} {\bibinfo {author} {\bibfnamefont {J.~D.}\ \bibnamefont
  {Bodyfelt}}, \bibinfo {author} {\bibfnamefont {M.~C.}\ \bibnamefont {Zheng}},
  \bibinfo {author} {\bibfnamefont {R.}~\bibnamefont {Fleischmann}}, \ and\
  \bibinfo {author} {\bibfnamefont {T.}~\bibnamefont {Kottos}},\ }\href@noop {}
  {\bibfield  {journal} {\bibinfo  {journal} {Phys. Rev. E}\ }\textbf {\bibinfo
  {volume} {87}},\ \bibinfo {pages} {020101} (\bibinfo {year}
  {2013})}\BibitemShut {NoStop}%
\bibitem [{\citenamefont {Anderson}(1958)}]{PWAnderson:PR58_Localization}%
  \BibitemOpen
  \bibfield  {author} {\bibinfo {author} {\bibfnamefont {P.~W.}\ \bibnamefont
  {Anderson}},\ }\href@noop {} {\bibfield  {journal} {\bibinfo  {journal}
  {Phys. Rev.}\ }\textbf {\bibinfo {volume} {109}},\ \bibinfo {pages} {1492}
  (\bibinfo {year} {1958})}\BibitemShut {NoStop}%
\bibitem [{\citenamefont {Hsiao}\ \emph {et~al.}(2013)\citenamefont {Hsiao},
  \citenamefont {Chang}, \citenamefont {Liou}, \citenamefont {Chu},
  \citenamefont {Lee},\ and\ \citenamefont
  {Chang}}]{TKHsiao:NatNano13_Observation}%
  \BibitemOpen
  \bibfield  {author} {\bibinfo {author} {\bibfnamefont {T.-K.}\ \bibnamefont
  {Hsiao}}, \bibinfo {author} {\bibfnamefont {H.-K.}\ \bibnamefont {Chang}},
  \bibinfo {author} {\bibfnamefont {S.-C.}\ \bibnamefont {Liou}}, \bibinfo
  {author} {\bibfnamefont {M.-W.}\ \bibnamefont {Chu}}, \bibinfo {author}
  {\bibfnamefont {S.-C.}\ \bibnamefont {Lee}}, \ and\ \bibinfo {author}
  {\bibfnamefont {C.-W.}\ \bibnamefont {Chang}},\ }\href@noop {} {\bibfield
  {journal} {\bibinfo  {journal} {Nat. Nanotech.}\ }\textbf {\bibinfo {volume}
  {8}},\ \bibinfo {pages} {534} (\bibinfo {year} {2013})}\BibitemShut {NoStop}%
\bibitem [{\citenamefont {Chang}\ \emph {et~al.}(2008)\citenamefont {Chang},
  \citenamefont {Okawa}, \citenamefont {Garcia}, \citenamefont {Majumdar},\
  and\ \citenamefont {Zettl}}]{CWChang:PRL08_Breakdown}%
  \BibitemOpen
  \bibfield  {author} {\bibinfo {author} {\bibfnamefont {C.~W.}\ \bibnamefont
  {Chang}}, \bibinfo {author} {\bibfnamefont {D.}~\bibnamefont {Okawa}},
  \bibinfo {author} {\bibfnamefont {H.}~\bibnamefont {Garcia}}, \bibinfo
  {author} {\bibfnamefont {A.}~\bibnamefont {Majumdar}}, \ and\ \bibinfo
  {author} {\bibfnamefont {A.}~\bibnamefont {Zettl}},\ }\href@noop {}
  {\bibfield  {journal} {\bibinfo  {journal} {Phys. Rev. Lett.}\ }\textbf
  {\bibinfo {volume} {101}},\ \bibinfo {pages} {075903} (\bibinfo {year}
  {2008})}\BibitemShut {NoStop}%
\bibitem [{\citenamefont {Savi\ifmmode~\acute{c}\else \'{c}\fi{}}\ \emph
  {et~al.}(2008)\citenamefont {Savi\ifmmode~\acute{c}\else \'{c}\fi{}},
  \citenamefont {Mingo},\ and\ \citenamefont
  {Stewart}}]{ISavic:PRL08_CNTLocalization}%
  \BibitemOpen
  \bibfield  {author} {\bibinfo {author} {\bibfnamefont {I.}~\bibnamefont
  {Savi\ifmmode~\acute{c}\else \'{c}\fi{}}}, \bibinfo {author} {\bibfnamefont
  {N.}~\bibnamefont {Mingo}}, \ and\ \bibinfo {author} {\bibfnamefont {D.~A.}\
  \bibnamefont {Stewart}},\ }\href@noop {} {\bibfield  {journal} {\bibinfo
  {journal} {Phys. Rev. Lett.}\ }\textbf {\bibinfo {volume} {101}},\ \bibinfo
  {pages} {165502} (\bibinfo {year} {2008})}\BibitemShut {NoStop}%
\bibitem [{\citenamefont {Yamamoto}\ \emph {et~al.}(2011)\citenamefont
  {Yamamoto}, \citenamefont {Sasaoka},\ and\ \citenamefont
  {Watanabe}}]{TYamamoto:PRL11_Universality}%
  \BibitemOpen
  \bibfield  {author} {\bibinfo {author} {\bibfnamefont {T.}~\bibnamefont
  {Yamamoto}}, \bibinfo {author} {\bibfnamefont {K.}~\bibnamefont {Sasaoka}}, \
  and\ \bibinfo {author} {\bibfnamefont {S.}~\bibnamefont {Watanabe}},\
  }\href@noop {} {\bibfield  {journal} {\bibinfo  {journal} {Phys. Rev. Lett.}\
  }\textbf {\bibinfo {volume} {106}},\ \bibinfo {pages} {215503} (\bibinfo
  {year} {2011})}\BibitemShut {NoStop}%
\bibitem [{\citenamefont {Kim}\ \emph {et~al.}(2001)\citenamefont {Kim},
  \citenamefont {Shi}, \citenamefont {Majumdar},\ and\ \citenamefont
  {McEuen}}]{PKim:PRL01_Thermal}%
  \BibitemOpen
  \bibfield  {author} {\bibinfo {author} {\bibfnamefont {P.}~\bibnamefont
  {Kim}}, \bibinfo {author} {\bibfnamefont {L.}~\bibnamefont {Shi}}, \bibinfo
  {author} {\bibfnamefont {A.}~\bibnamefont {Majumdar}}, \ and\ \bibinfo
  {author} {\bibfnamefont {P.~L.}\ \bibnamefont {McEuen}},\ }\href@noop {}
  {\bibfield  {journal} {\bibinfo  {journal} {Phys. Rev. Lett.}\ }\textbf
  {\bibinfo {volume} {87}},\ \bibinfo {pages} {215502} (\bibinfo {year}
  {2001})}\BibitemShut {NoStop}%
\bibitem [{\citenamefont {Fujii}\ \emph {et~al.}(2005)\citenamefont {Fujii},
  \citenamefont {Zhang}, \citenamefont {Xie}, \citenamefont {Ago},
  \citenamefont {Takahashi}, \citenamefont {Ikuta}, \citenamefont {Abe},\ and\
  \citenamefont {Shimizu}}]{MFujii:PRL05_Measuring}%
  \BibitemOpen
  \bibfield  {author} {\bibinfo {author} {\bibfnamefont {M.}~\bibnamefont
  {Fujii}}, \bibinfo {author} {\bibfnamefont {X.}~\bibnamefont {Zhang}},
  \bibinfo {author} {\bibfnamefont {H.}~\bibnamefont {Xie}}, \bibinfo {author}
  {\bibfnamefont {H.}~\bibnamefont {Ago}}, \bibinfo {author} {\bibfnamefont
  {K.}~\bibnamefont {Takahashi}}, \bibinfo {author} {\bibfnamefont
  {T.}~\bibnamefont {Ikuta}}, \bibinfo {author} {\bibfnamefont
  {H.}~\bibnamefont {Abe}}, \ and\ \bibinfo {author} {\bibfnamefont
  {T.}~\bibnamefont {Shimizu}},\ }\href@noop {} {\bibfield  {journal} {\bibinfo
   {journal} {Phys. Rev. Lett.}\ }\textbf {\bibinfo {volume} {95}},\ \bibinfo
  {pages} {065502} (\bibinfo {year} {2005})}\BibitemShut {NoStop}%
\bibitem [{\citenamefont {Hsu}\ \emph {et~al.}(2008)\citenamefont {Hsu},
  \citenamefont {Kumar}, \citenamefont {Bushmaker}, \citenamefont {Cronin},
  \citenamefont {Pettes}, \citenamefont {Shi}, \citenamefont {Brintlinger},
  \citenamefont {Fuhrer},\ and\ \citenamefont {Cumings}}]{IHsu:APL08}%
  \BibitemOpen
  \bibfield  {author} {\bibinfo {author} {\bibfnamefont {I.-K.}\ \bibnamefont
  {Hsu}}, \bibinfo {author} {\bibfnamefont {R.}~\bibnamefont {Kumar}}, \bibinfo
  {author} {\bibfnamefont {A.}~\bibnamefont {Bushmaker}}, \bibinfo {author}
  {\bibfnamefont {S.~B.}\ \bibnamefont {Cronin}}, \bibinfo {author}
  {\bibfnamefont {M.~T.}\ \bibnamefont {Pettes}}, \bibinfo {author}
  {\bibfnamefont {L.}~\bibnamefont {Shi}}, \bibinfo {author} {\bibfnamefont
  {T.}~\bibnamefont {Brintlinger}}, \bibinfo {author} {\bibfnamefont {M.~S.}\
  \bibnamefont {Fuhrer}}, \ and\ \bibinfo {author} {\bibfnamefont
  {J.}~\bibnamefont {Cumings}},\ }\href@noop {} {\bibfield  {journal} {\bibinfo
   {journal} {Applied Physics Letters}\ }\textbf {\bibinfo {volume} {92}},\
  \bibinfo {pages} {063119} (\bibinfo {year} {2008})}\BibitemShut {NoStop}%
\bibitem [{\citenamefont {Dyson}(1953)}]{FJDyson:PR53_DisorderedLinearChain}%
  \BibitemOpen
  \bibfield  {author} {\bibinfo {author} {\bibfnamefont {F.~J.}\ \bibnamefont
  {Dyson}},\ }\href@noop {} {\bibfield  {journal} {\bibinfo  {journal} {Phys.
  Rev.}\ }\textbf {\bibinfo {volume} {92}},\ \bibinfo {pages} {1331} (\bibinfo
  {year} {1953})}\BibitemShut {NoStop}%
\bibitem [{\citenamefont {Zhang}\ \emph {et~al.}(2007)\citenamefont {Zhang},
  \citenamefont {Fisher},\ and\ \citenamefont {Mingo}}]{WZhang:NHT07}%
  \BibitemOpen
  \bibfield  {author} {\bibinfo {author} {\bibfnamefont {W.}~\bibnamefont
  {Zhang}}, \bibinfo {author} {\bibfnamefont {T.}~\bibnamefont {Fisher}}, \
  and\ \bibinfo {author} {\bibfnamefont {N.}~\bibnamefont {Mingo}},\
  }\href@noop {} {\bibfield  {journal} {\bibinfo  {journal} {Numerical Heat
  Transfer, Part B: Fundamentals}\ }\textbf {\bibinfo {volume} {51}},\ \bibinfo
  {pages} {333} (\bibinfo {year} {2007})}\BibitemShut {NoStop}%
\bibitem [{\citenamefont {Mingo}(2009)}]{NMingo:Springer09}%
  \BibitemOpen
  \bibfield  {author} {\bibinfo {author} {\bibfnamefont {N.}~\bibnamefont
  {Mingo}},\ }in\ \href@noop {} {\emph {\bibinfo {booktitle} {Thermal
  Nanosystems and Nanomaterials}}}\ (\bibinfo  {publisher} {Springer},\
  \bibinfo {year} {2009})\ pp.\ \bibinfo {pages} {63--94}\BibitemShut {NoStop}%
\bibitem [{\citenamefont {Caroli}\ \emph {et~al.}(1971)\citenamefont {Caroli},
  \citenamefont {Combescot}, \citenamefont {Nozieres},\ and\ \citenamefont
  {Saint-James}}]{CCaroli:JPhysC71}%
  \BibitemOpen
  \bibfield  {author} {\bibinfo {author} {\bibfnamefont {C.}~\bibnamefont
  {Caroli}}, \bibinfo {author} {\bibfnamefont {R.}~\bibnamefont {Combescot}},
  \bibinfo {author} {\bibfnamefont {P.}~\bibnamefont {Nozieres}}, \ and\
  \bibinfo {author} {\bibfnamefont {D.}~\bibnamefont {Saint-James}},\
  }\href@noop {} {\bibfield  {journal} {\bibinfo  {journal} {J. Phys. C}\
  }\textbf {\bibinfo {volume} {4}},\ \bibinfo {pages} {916} (\bibinfo {year}
  {1971})}\BibitemShut {NoStop}%
\bibitem [{\citenamefont {Das}\ and\ \citenamefont
  {Dhar}(2012)}]{SGDas:EPJB12_Landauer}%
  \BibitemOpen
  \bibfield  {author} {\bibinfo {author} {\bibfnamefont {S.~G.}\ \bibnamefont
  {Das}}\ and\ \bibinfo {author} {\bibfnamefont {A.}~\bibnamefont {Dhar}},\
  }\href@noop {} {\bibfield  {journal} {\bibinfo  {journal} {Eur. Phys. J. B}\
  }\textbf {\bibinfo {volume} {11}},\ \bibinfo {pages} {1857} (\bibinfo {year}
  {2012})}\BibitemShut {NoStop}%
\bibitem [{\citenamefont {Holland}(1963)}]{MGHolland:PR63_Analysis}%
  \BibitemOpen
  \bibfield  {author} {\bibinfo {author} {\bibfnamefont {M.~G.}\ \bibnamefont
  {Holland}},\ }\href@noop {} {\bibfield  {journal} {\bibinfo  {journal} {Phys.
  Rev.}\ }\textbf {\bibinfo {volume} {132}},\ \bibinfo {pages} {2461} (\bibinfo
  {year} {1963})}\BibitemShut {NoStop}%
\end{thebibliography}%

\end{document}